\documentclass[reprint,
nofootinbib,
amsmath,
amssymb,
aps,
pre,
longbibliography
]{revtex4-2}

\usepackage{physics}
\usepackage{soul}
\usepackage{tabularx}
\usepackage{xcolor}
\usepackage{comment}
\usepackage{float}

\usepackage{placeins}

\usepackage{graphicx}% Include figure files
\usepackage{dcolumn}% Align table columns on decimal point
\usepackage{bm}% bold math
\newcommand{\PE}{\mathrm{PE}}

\newcommand{\Str}{\mathrm{Str}\,}

\begin{document}

\title{Supergroup Invariants and the Brane/Negative Brane Expansion}


\author{Kasia Budzik}
\email{kbudzik@fas.harvard.edu}
\affiliation{%
Center for the Fundamental Laws of Nature, Harvard University, Cambridge, MA 02138, USA}%

\begin{abstract}
We propose a Molien--Weyl-type formula computing generating functions of invariants of supergroups $U(N|M)$, i.e. polynomials in supertraces, which arise as gauge groups of brane/negative brane systems in string theory. We either prove or numerically verify the formula in various examples. The formula further leads to a new expansion relating finite-$N$ and infinite-$N$ indices of $U(N)$ gauge theories. We comment on its relation to Murthy's Giant Graviton expansion, for which we suggest a physical interpretation in terms of ``Koszul dual" branes and negative branes.
\end{abstract}

\maketitle

%\tableofcontents

\section{Introduction and Main Results}

Supergroups $U(N|M)$ appear as gauge groups in brane/negative brane systems in superstring theory \cite{Okuda:2006fb, Vafa:2014iua, Mikhaylov:2014aoa,Dijkgraaf:2016lym}, topological strings\footnote{In the topological strings context they are also sometimes referred to as anti-branes \cite{Vafa:2001qf}, although they are not related to anti-branes in superstring theory, which have positive tension.} \cite{Vafa:2001qf}, and more recently in the context of the Giant Graviton expansion \cite{Murthy:2022ien, Liu:2022olj, Eniceicu:2023uvd}. For a recent overview of supergroup gauge theories, see \cite{Kimura:2023iup}.

In this note, we study Hilbert series (generating functions) of $U(N|M)$ invariants\footnote{These arise in physics e.g. as free partition functions of $U(N|M)$ gauge theories or supersymmetric indices.} from which we derive identities connecting finite- and infinite-$N$ Hilbert series of $U(N)$ invariants.

Recall that for the ordinary $U(N)$ groups, the generating function of gauge-invariant polynomials in adjoint-valued fields is computed by the Molien--Weyl formula:
\begin{align}
    Z_N  \! = \! \frac{1}{N!} \oint_{|u_i|=1} \prod_{i=1}^N \frac{\dd u_i}{2\pi i u_i} \Delta(u) \, \PE\qty[ \qty(\sum_{i=1}^N u_i\!)\qty(\sum_{i=1}^N u_i^{-1}\!) f ] , \label{eq:UN}
\end{align}
where $f$ is the single-letter index encoding contributions of adjoint fields, $\PE$ denotes the plethystic exponential, and
\begin{align}
    \Delta(u) = \prod_{i\neq j}^N \qty( 1-\frac{u_i}{u_j} ) 
\end{align}
is the Vandermonde determinant. %Such integrals arise in physics e.g. as free partition functions of $U(N)$ gauge theories or as supersymmetric indices. %Hilbert series of invariants of $U(N)$ matrices has been studied e.g. in ..........

In the $U(N|M)$ case, the basic gauge-invariant operators are polynomials in supertraces, defined by
\begin{align}
    \Str X = \Tr A - \Tr D \, , \quad X=\mqty( A & B \\ C & D) \, ,
\end{align}
where $A$ and $D$ are $N\times N$ and $M\times M$ blocks with complex bosonic entries, and $B$ and $C$ are $N\times M$ and $M\times N$ blocks with complex Grassmann entries.

We propose the identity\footnote{$Z_N$ can be included in the sum as $Z^+_{N|0}$.}
\begin{align}
    Z_{\infty} = Z_N + \sum_{k=1}^\infty Z^+_{N+k|k} \, , \label{eq:inf}
\end{align}
where $Z^+_{N+k|k}$ are Molien--Weyl-type integrals:\footnote{In Section \ref{sec:vectors}, we consider also (anti)fundamental degrees of freedom.}
\begin{align}
    Z^+_{N|M} = \frac{1}{N!M!} \oint_{\substack{|u_i|=1-\varepsilon \\ |v_j|=1+\varepsilon}} \prod_{i=1}^N \frac{\dd u_i}{2\pi i u_i} \prod_{j=1}^M \frac{\dd v_j}{2\pi i v_j} \Delta(u,v) \nonumber \\ 
    \PE\qty[ \qty(\sum_{i=1}^N u_i - \sum_{j=1}^M v_j ) \qty(\sum_{i=1}^N u_i^{-1} - \sum_{j=1}^M v_j^{-1} ) f ] \, , \label{eq:Zplus}
\end{align}
and
\begin{align}
    \Delta(u,v) = \frac{\prod_{i\neq j}^N \qty(1-\frac{u_i}{u_j}) \prod_{i\neq j}^M \qty(1-\frac{v_i}{v_j})}{\prod_{i=1}^N \prod_{j=1}^M \qty(1-\frac{u_i}{v_j}) \qty(1-\frac{v_j}{u_i})} \,  \label{eq:van}
\end{align}
is the super-Vandermonde determinant (Berezinian).

We conjecture that the partial sum
\begin{align}
    Z_{N+K|K} = \sum_{k=0}^K Z^+_{N+k|k} \, ,\label{eq:main}
\end{align}
computes the Hilbert series of $U(N\!+\!K|K)$ invariants, i.e. polynomials in supertraces.

In the $K\rightarrow\infty$ limit, (super)trace relations are absent and so the expansion \eqref{eq:inf} can be understood as the equality
\begin{align}
    Z_\infty = Z_{N+\infty|\infty} \,  . \label{eq:inf0}
\end{align}

The integrals \eqref{eq:Zplus} generalize the standard Molien--Weyl formula \eqref{eq:UN} to $U(N|M)$. The super-Vandermonde \eqref{eq:van} has a pole on the standard unit contour. We find that after deforming the contour, \eqref{eq:Zplus} produces only part of the result, and the total Hilbert series is obtained by summing the MW integrals as in \eqref{eq:main}. Integrals of the form \eqref{eq:Zplus} were also studied in \cite{berele2010computing}, where it was likewise found that they capture part of the full Hilbert series of superinvariants. It would be interesting to find a single integral representation with a contour/residue prescription that reproduces this sum; we leave this problem for future work.

For the one matrix case, i.e. $f=x$, we prove the expansion \eqref{eq:inf} and verify that the partial sums \eqref{eq:main} reproduce the Hilbert series of supersymmetric polynomials \cite{Molev:2015fza, sergeev2004deformed}, which coincide with the ring of superinvariants of a single matrix. See Section \ref{sec:onematrix} and Appendix \ref{app:integrals}.

We further test the expansion \eqref{eq:inf} and the conjectured Hilbert series \eqref{eq:main} of superinvariants in more examples in the supplemental material.

For the two matrix case $f=x+y$ and the three matrix case $f=x+y+z$, for which the Hilbert series are not in general known, we compute the partial sums \eqref{eq:main} for select small values of $N,K$. We verify them by explicitly enumerating superinvariants up to length six. See Sections \ref{app:2} and \ref{app:3}.

We further verify the expansion \eqref{eq:inf} in these cases in Section \ref{app:tables}. We also consider BPS subsectors of $\mathcal{N}=4$ SYM, including the BMN truncation of $\mathcal{N}=4$ SYM with $f=1-(1-x)(1-y)(1-z)$. For fermionic vector degrees of freedom, the expansion \eqref{eq:inf} leads to interesting identities, see Sections \ref{sec:vectors} and \ref{app:vectors}.

\begin{comment}
    For the two matrix case $f=x+y$ and the three matrix case $f=x+y+z$, for which the Hilbert series are not in general known, we compute the partial sums \eqref{eq:main} for select small values of $N+K,K$. We verify them by explicitly enumerating superinvariants up to length six. See Appendices \ref{app:2} and \ref{app:3}.

    We further test the expansion \eqref{eq:inf} in these cases in Appendix \ref{app:tables}. We also consider BPS subsectors of $\mathcal{N}=4$ SYM, including the BMN truncation of $\mathcal{N}=4$ SYM with $f=1-(1-x)(1-y)(1-z)$. For fermionic vector degrees of freedom, the expansion leads to interesting identities, see Section \ref{sec:vectors} and Appendix \ref{app:vectors}.
\end{comment}

The identity \eqref{eq:inf} looks similar to the Giant Graviton expansion (GGE) \cite{Imamura:2021ytr,Gaiotto:2021xce,Murthy:2022ien}. For $N=0$, it reduces to Murthy's GGE \cite{Murthy:2022ien, Liu:2022olj, Eniceicu:2023uvd} (but for transverse branes), which we show in Section \ref{sec:GGE}.  The general $N$ case will be studied in a subsequent paper \cite{inprogress}. We further comment on the physical interpretation of Murthy's GGE in terms of indices of ``Koszul dual" branes and negative branes in Section \ref{sec:GGE} and Appendix \ref{app:branes}.

%[AAA] We hope identity might lead to interesting combinatorial identities and progress in studying Hilbert series of supergroup invariants. Final message is that negative branes are important for finite $N$!

\section{One matrix}
\label{sec:onematrix}

For $f=x$, the integrals \eqref{eq:Zplus} can be evaluated exactly by residues using methods from \cite{Gadde:2025yoa}, see Appendix \ref{app:integrals}. The result takes the form
\begin{align}
    Z^+_{N|M} = \frac{x^{NM}}{\prod_{n=1}^{N} (1-x^n) \prod_{m=1}^{M} (1-x^m) } \, . \label{eq:Zp1m}
\end{align}
Notice that the ``corrections" start at order $\mathcal{O}(x^{NM})$.

In this simple case, we can show that the infinite sum \eqref{eq:inf} reproduces known combinatorial identities, and the partial sum \eqref{eq:main} agrees with the Hilbert series of supersymmetric polynomials \cite{Molev:2015fza, sergeev2004deformed}, which coincide with the superinvariants of one matrix.

Analogous partial results for the Hilbert series of two and three matrices are given in Sections~\ref{app:2} and~\ref{app:3}, and for the expansion \eqref{eq:inf} in Section \ref{app:tables} in the supplementary material.

\subsection{Expansion}

For one matrix, the identity \eqref{eq:inf} can be written as
\begin{align}
    \frac{1}{(x;x)_\infty} = \sum_{k=0}^\infty \frac{x^{(N+k)k}}{(x;x)_{N+k}(x;x)_k} \, , \label{eq:Cauchy}
\end{align}
where $(x;x)_n=(1-x)(1-x^2)\dots(1-x^n)$. This is a corollary of a combinatorial identity due to Cauchy \cite[Eq.~(2.2.8)]{Andrews_1984}, see Appendix \ref{app:Cauchy}. For $N=0$, it reduces to an identity\footnote{The combinatorial proof can be found e.g. in \cite[p.~27--28]{Andrews_1984}.} by Euler \cite[Eq.~(2.2.9)]{Andrews_1984}:
\begin{align}
    \frac{1}{(x;x)_\infty} = \sum_{k=0}^\infty \frac{x^{k^2}}{(x;x)_k^2} \, .
\end{align}

\subsection{$U(N|M)$ invariants}

Denote the eigenvalues of $X\in \mathfrak{u}(N|M)$ as $z_i$, $i=1,\dots,N$ and $w_j$, $j=1,\dots,M$. The basic gauge-invariant operators are the supertraces:
\begin{align}
    \Str X^n = \sum_{i=1}^{N} z_i^n - \sum_{j=1}^M w_j^n \, , \quad n\in\mathbb{Z}_{>0} \, . \label{eq:polys}
\end{align}
The first $N+M$ supertraces are algebraically independent, while those with $n>N+M$ can be expressed as rational functions of those with $n\leq N+M$.\footnote{This is unlike the $U(N)$ case where traces $\Tr X^n$ with $n>N$ are polynomial combinations of traces with $n\leq N$. Therefore the polynomial ring of supertraces is not finitely generated \cite{sergeev1999invariant}.} 

The ring generated by \eqref{eq:polys} is known as the ring of supersymmetric polynomials \cite{STEMBRIDGE1985439,sergeev1999invariant}. These are functions $f(z_i,w_j)$ that are invariant under separate permutations of $z_i$ and $w_j$, and satisfy the condition 
\begin{align}
    \pdv{f}{z_i} + \pdv{f}{w_j} \Big|_{z_i=w_j} = 0 \, .
\end{align}

The Hilbert series of supersymmetric polynomials can be computed by counting Young diagrams that fit inside an $(N,M)$-hook \cite{Berele:1987yi, Fulton_Pragacz_1998}. The Hilbert series matching our conjecture \eqref{eq:main} was given in \cite{Molev:2015fza}:
\begin{align}
    Z_{N|M} = \sum_{k=0}^{\min(N,M)} \frac{x^{(N-k)(M-k)}}{(x;x)_{N-k}(x;x)_{M-k}} \, . \label{eq:mine}
\end{align}
Another formula was obtained in \cite{sergeev2004deformed}:\footnote{ $\widetilde{Z}_{N|M} = P_{N,M}$ in the notation of \cite{sergeev2004deformed}. $\tilde{Z}_{N|M}$ does not manifestly look symmetric in $N$ and $M$ and $\tilde{Z}_{N|M}=\tilde{Z}_{M|N}$ is another (unproven) combinatorial identity.}\footnote{Another recursive formula was found in \cite{orellana2000some}.}
\begin{align}
    \widetilde{Z}_{N|M} = \frac{1}{(x;x)_{N}} \sum_{k=0}^M \frac{x^{(N+1)k}}{(x;x)_k} \, . \label{eq:theirs}
\end{align}

The summands in \eqref{eq:mine} and \eqref{eq:theirs} are different but they add up to the same answer in all examples checked, e.g.:
\begin{align}
     Z_{2|1} &= \frac{1}{1-x}+\frac{x^2}{(1-x)^2 (1-x^2)} = \frac{1-x+x^3}{(1-x)^2(1-x^2)}  \\
    \widetilde{Z}_{2|1} &= \frac{1}{(1\!-\!x)(1\!-\!x^2)}\! +\! \frac{x^3}{(1\!-\!x)^2 (1\!-\!x^2)}\! =\! \frac{1-x+x^3}{(1\!-\!x)^2(1\!-\!x^2)} \, .
\end{align}
One can also check for wide ranges of $N$ and $M$ that they agree numerically (after substituting e.g. $x=0.99$). %e.g. after substituting $x=0.99$:
%\begin{align}
%    Z_{95|42} &\sim 1.18256 \times 10^{69} \\
%    \widetilde{Z}_{95|42} &\sim 1.18256 \times 10^{69} \\
%    Z_{95|42} &\sim 1.18256 \times 10^{69} \\
%    \widetilde{Z}_{95|42} &\sim 1.18256 \times 10^{69} \, .
%\end{align}

Let us consider a special example.

\paragraph{$U(1|1)$}

The formula \eqref{eq:mine} gives
\begin{align}
    Z_{1|1} = 1 + \frac{x}{(1-x)^2} \, .
\end{align}

Let us recover this from counting supersymmetric polynomials, following \cite{brookner2016cohen}. The $U(1|1)$ generators are
\begin{align}
    z_1^n - w_1^n \, , \quad n\in\mathbb{Z}_{>0} \, .
\end{align}
Change the variables to
\begin{align}
    p= z_1-w_1 \, , \quad q=z_1+w_1 \, .
\end{align}
Then one can find a basis of the generators
\begin{align}
    p, pq, p q^2, p q^3, \dots \, .
\end{align}
But these generate all the polynomials in two variables $p$ and $q$, without monomials in $q$, i.e. $\mathbb{C}\oplus p\, \mathbb{C}[p,q]$, whose Hilbert series is
\begin{align}
    1 + \frac{x}{(1-x)^2} \, .
\end{align}

\section{Vectors}
\label{sec:vectors}

One can extend the integrals \eqref{eq:Zplus} to include vector degrees of freedom by adding to the arguments of the $\PE$ the terms
\begin{align}
       \qty(\sum_{i=1}^N u_i - \sum_{j=1}^M v_j ) \bar{h} + \qty(\sum_{i=1}^N u_i^{-1} - \sum_{j=1}^M v_j^{-1} ) h \, , \label{eq:hbarh}
\end{align}
where $\bar{h}$ and $h$ are single-letter indices encoding (anti)fundamental contributions.

In this section, we conjecture the form of the expansion \eqref{eq:inf} for fermionic vectors, in which case it truncates to a finite number of terms. We present evidence in Section \ref{app:vectors} in the supplemental material.

\subsection{$(\bar{F},F)=(1,1)$}

First, consider the simplest case of one anti-fundamental and one fundamental fermion, $\bar{\theta}$ and $\theta$, with fugacities $\bar{h}=-\bar{s}$ and $h=-s$. The gauge-invariant combination is the bilinear $\bar{\theta}\theta$.

%\begin{align}
%    \bar{\psi} = \mqty( \bar{\theta} & \bar{x} ) \in (\mathbb{C}^{N|M})^* \, , \quad \psi = \mqty( \theta \\ x ) \in \mathbb{C}^{N|M}  \, ,
%\end{align}
%where $\bar{\theta},\theta$ are odd vectors and $\bar{x},x$ bosonic vectors, with fugacities $\bar{h}=-\bar{s}$ and $h=s$. The gauge-invariant combination is the bilinear $\bar{\psi}\psi$.

%$\bar{\theta}$ and $\theta$, with fugacities $\bar{h}=-\bar{s}$ and $h=s$. The gauge-invariant combination is $\bar{\psi}\psi$.

The $U(N)$ partition function is
\begin{align}
    Z_N &= 1+ \bar{s} s + \dots + (\bar{s} s)^N = \frac{1-(\bar{s}s)^{N+1} }{1-\bar{s} s}
\end{align}
since $(\bar{\theta}\theta)^{N+1}=0$ and the infinite-$N$ limit is
\begin{align}
    Z_\infty &= \frac{1}{1-\bar{s} s} \, .
\end{align}

The integrals \eqref{eq:Zplus} including \eqref{eq:hbarh} take the form
\begin{gather}
Z_{N|M}^+ = \frac{1}{N! M!} \oint_{\substack{|u_i|=1-\varepsilon \\ |v_j|=1+\varepsilon}} \prod_{i=1}^N \frac{\dd u_i}{2\pi i u_i} \prod_{j=1}^M \frac{\dd v_j}{2\pi i v_j} \nonumber\\
   \Delta(u,v) \frac{\prod_{i=1}^N\qty(1-u_i \bar{s})\qty(1- \frac{s}{u_i})}{\prod_{j=1}^M\qty(1-v_j\bar{s})\qty(1-\frac{s}{v_j})} \, . \label{eq:ZPlusvector}
\end{gather}
In Section \ref{app:vectors}, we show evidence that
\begin{align}
    Z_{N+1|1}^+ &= \frac{(\bar{s}s)^{N+1}}{1-\bar{s} s}
\end{align}
and all the other integrals vanish:
\begin{align}
    Z_{N+k|k}^+ = 0 \, , \quad k\geq 2 \, .
\end{align}

Therefore, we conjecture that the expansion \eqref{eq:inf} for $(\bar{F},F)=(1,1)$ reduces to just
\begin{align}
    Z_\infty = Z_N + Z_{N+1|1}^+ \, . \label{eq:FbarF11}
\end{align}

Since the last components of $\bar\theta\in (\mathbb{C}^{N+1|1})^*$ and $\theta\in \mathbb{C}^{N+1|1}$ are even, $(\bar\theta\theta)^{N+1} \neq 0$ and the trace relation is lifted, in agreement with \eqref{eq:FbarF11}. 

\subsection{General $(\bar{F},F)$}

Now, consider $\bar{F}$ anti-fundamentals $\bar{\theta}_I$ and $F$ fundamentals $\theta_J$, contributing $\bar{h}=-\sum_I \bar{s}_I$ and $h=-\sum_J s_J$. The gauge-invariant operators are generated by the bilinears $\bar{\theta}_I\theta_J$.

The $U(N)$ partition function is a finite sum of products of binomials $\bar{s}_I s_J$ such that a power of a single variable does not exceed $N$ due to trace relations, while the infinite-$N$ limit is
\begin{align}
    Z_\infty = \prod_{I=1}^{\bar{F}} \prod_{J=1}^F \frac{1}{1-\bar{s}_I s_J} \, . \label{eq:vectorsinf}
\end{align}

We conjecture that the expansion \eqref{eq:inf} truncates:
\begin{align}
    Z_\infty = \sum_{k=0}^{\min(\bar{F},F)} Z^+_{N+k|k} \, . \label{eq:FbarF}
\end{align}
and
\begin{align}
    Z_{N+k|k}^+ = 0 \, , \quad k>\min(\bar{F},F) \, ,
\end{align}
which we verify in Section \ref{app:vectors}.

\section{Comparison with GGE}
\label{sec:GGE}

The identity \eqref{eq:inf} can be seen as a series of corrections\footnote{Although the corrections $Z^+_{N+k|k}$ require $N+2k$ integrals, compared to $N$ integrals for $Z_N$, so \eqref{eq:expansion} is not advantageous for computing finite-$N$ indices.} from the infinite-$N$ index to the finite-$N$ index:
\begin{align}
    Z_N = Z_\infty - \sum_{k=1}^\infty Z^+_{N+k|k} \, . \label{eq:expansion}
\end{align}

This formula looks similar to a Giant Graviton expansion (GGE) \cite{Imamura:2021ytr,Gaiotto:2021xce,Murthy:2022ien}, in which the finite-$N$ corrections are related to indices of the Giant Graviton branes in the holographic dual.\footnote{One can check that $-Z^+_{N+k|k}$ differ from $x^{kN}Z_{\infty}\hat{Z}_k$ corrections of \cite{Gaiotto:2021xce} and $Z_{\infty}G^{(k)}_N$ corrections of \cite{Murthy:2022ien}.} Let us in particular consider the GGE derived by Murthy \cite{Murthy:2022ien}:
\begin{align}
    Z_N = Z_\infty \sum_{k=0}^\infty G^{(k)}_N \, . \label{eq:MGGE}
\end{align}
It was shown by Eniceicu \cite{Eniceicu:2023uvd} (see also \cite{Liu:2022olj}), that $G^{(k)}_N$ admit a representation as $U(k|k)$ matrix integrals:
\begin{gather}
    G^{(k)}_N = \frac{1}{(k!)^2}\oint_{\substack{|u_i|=1-\varepsilon \\ |v_i|=1+\varepsilon}} \prod_{i=1}^k \qty[\frac{\dd u_i}{2\pi i u_i}\frac{\dd v_i}{2\pi i v_i} \qty(\frac{u_i}{v_i})^N ] \Delta(u,v) \nonumber \\ 
    \qquad \PE \qty[ \sum_{i=1}^k (u_i - v_i ) \sum_{i=1}^k (u_i^{-1} - v_i^{-1} ) \frac{-f}{1-f} ] \, . \label{eq:GkN}
\end{gather}

First, we remark that the corrections $G^{(k)}_N$ employ the single-letter index\footnote{Our $f^!=-\hat{f}$ from \cite{Murthy:2022ien}.}
\begin{align}
    f^! \equiv - \frac{f}{1-f} \, .
\end{align}
This is the single-letter index of probe branes completely transverse to the original stack described by $f$, aka ``Koszul dual" branes,\footnote{We choose this name because the worldvolume operator algebras of fully transverse branes are Koszul dual, i.e. $A^!=(\mathbb{C}\otimes_A\mathbb{C})^\vee$, and their characters $\chi(A)=1-f$, $\chi(A^!)=1-f^!$ are precisely related by $(1-f)(1-f^!)=1$ \cite{Gaiotto:2024dwr}. Notice also $f^{!!}=f$.} rather than the standard Giant Graviton branes, which are (complex) codimension one with respect to the stack.\footnote{E.g. in case of $\mathcal{N}=4$ SYM, these are D5-branes, (which will be studied in \cite{inprogress2}) and not the standard Giant Graviton D3-branes.} See Appendix \ref{app:branes} for some examples. 

Then, notice that
\begin{align}
    Z^+_{k|k} [f^!] = G^{(k)}_{N=0}[f]
\end{align}
and\footnote{We thank Davide Gaiotto for pointing this out.}
\begin{align}
    Z_{\infty}[f^!] = \frac{1}{Z_{\infty}[f]} \, .
\end{align}
Therefore, the expansion \eqref{eq:expansion} for $N=0$,
\begin{align}
    Z_{\infty}[f] = \sum_{k=0}^\infty Z_{k|k}^+[f] \, ,
\end{align}
gives
\begin{align}
    1 = Z_\infty [f^!] \sum_{k=0}^\infty G^{(k)}_{N=0} [f^!] \, ,
\end{align}
which is the Murthy's GGE for $N=0$ and $f\mapsto f^!$. The $k|k$ branes are now identified as the Koszul dual branes of Murthy's GGE. This will be explored at the derived level and for general $N$ in the upcoming work \cite{inprogress}.

We conclude with the observation that the partial sum $\sum^k_j G^{(j)}_{N}$ (rather than a single term $G^{(k)}_{N}$) in Murthy's GGE should be identified with the index of $k|k$ Koszul dual branes (in the presence of the original stack of $N$ branes).

\begin{acknowledgments}
The author would like to thank Allan Berele, Kevin Costello, Pavel Etingof, Davide Gaiotto, Elliott Gesteau, Matthew Heydeman, Andrew Strominger, Cumrun Vafa and Pedro Vieira for useful discussions and comments on the draft. This work was supported by the Simons Collaboration on Celestial Holography. Part of this work was performed at Aspen Center for Physics, which is supported by National Science Foundation grant PHY-2210452.
\end{acknowledgments}

\begin{appendix}

\section{Integrals for one matrix}
\label{app:integrals}

Let us evaluate the integral \eqref{eq:Zplus} for the one matrix case, i.e. $f=x$:
\begin{gather}
    Z^+_{N|M} = \frac{1}{N!M!} \frac{1}{(1-x)^{N+M}} \oint_{\substack{|u_i|=1-\varepsilon \\ |v_j|=1+\varepsilon}} \prod_{i=1}^N \frac{\dd u_i}{2\pi i u_i} \prod_{j=1}^M \frac{\dd v_j}{2\pi i v_j} \nonumber \\
    \Delta(u,v) \frac{\prod_{i=1}^N \prod_{j=1}^M \qty(1-\frac{u_i}{v_j}x)\qty(1-\frac{v_j}{u_i}x)}{\prod_{i\neq j}^N\qty(1-\frac{u_i}{u_j}x) \prod_{i\neq j}^M \qty(1-\frac{v_i}{v_j}x)} \, 
\end{gather}
where 
\begin{align}
    \Delta(u,v) = \frac{\prod_{i\neq j}^N \qty(1-\frac{u_i}{u_j}) \prod_{i\neq j}^M \qty(1-\frac{v_i}{v_j})}{\prod_{i=1}^N \prod_{j=1}^M \qty(1-\frac{u_i}{v_j}) \qty(1-\frac{v_j}{u_i})} \, .
\end{align}
We flip the $v_j$ contours, which contributes a factor $(-1)^M$.

We follow \cite{Gadde:2025yoa} to compute this integral by residues. First, we shift the poles $u_i=0$ and $v_j=\infty$ to $u_i=\sigma$ and $v_j=1/\sigma$, by
\begin{align}
    \frac{\dd u_i}{u_i} \frac{\dd v_j}{v_j} \mapsto \frac{\dd u_i}{u_i-\sigma} \frac{\dd v_j}{v_j-\sigma v_j^2} \, .
\end{align}
Unlike in \cite{Gadde:2025yoa}, the result now depends on $\sigma$, which we send to $0$ later. 

Note that residues with repeated poles $u_i^* = u_j^*$ for $i\neq j$ vanish due to the numerator factors. Thus there are $N!$ ordered ``tuples" of poles $(u_1^*,\dots,u_N^*)$ and analogously $M!$ tuples $(v_1^*,\dots,v_M^*)$. Due to the permutation symmetry of the integral they all contribute equally, so we can choose 
\begin{align}
    (u_1^*,u_2^*,u_3^*,\dots,u_N^*) &= (\sigma, \sigma x, \dots, \sigma x^{N-1}) \nonumber \\
    (v_1^*,v_2^*,v_3^*,\dots,v_M^*) &= \qty(\tfrac{1}{\sigma}, \tfrac{1}{\sigma x} , \dots , \tfrac{1}{\sigma x^{M-1}} ) \, . \label{eq:tuples}
\end{align}
Let $r_{N|M}$ denote the residue evaluated at these poles. Then
\begin{align}
    Z^+_{N|M} &= N! M! \, r_{N|M} \, .
\end{align}
We want to compute this recursively. The residues satisfy
\begin{align}
    r_{N|M} &= r_{N-1|M} \frac{1}{N} \frac{x^M}{1-x^N} \frac{(1-\sigma^2 x^{N+M-1})(1-\sigma^2 x^{N-2})}{(1-\sigma^2 x^{N-1})(1-\sigma^2x^{N+M-2})} \, ,
\end{align}
which in the limit $\sigma\rightarrow 0$ reduces to
\begin{align}
    r_{N|M} = r_{N-1|M} \frac{1}{N} \frac{x^M}{1-x^N}  \, . 
\end{align}
Using
\begin{align}
    Z^+_{N-1|M} &= (N-1)! M! \, r_{N-1|M} \, ,
\end{align}
we obtain recursion
\begin{align}
    Z^+_{N|M} &= Z^+_{N-1|M} \frac{x^M}{(1-x^N)} \, . \label{eq:rec}
\end{align}
Iterating \eqref{eq:rec} gives
\begin{align}
    Z^+_{N|M} = Z^+_{0|M} \frac{x^{NM}}{\prod_{n=1}^N(1-x^n)} \, .
\end{align}
Since
\begin{align}
    Z^+_{0|M} = Z_M = \frac{1}{\prod_{m=1}^M(1-x^m)} \, ,
\end{align}
we arrive at
\begin{align}
    Z^+_{N|M} = \frac{x^{NM}}{\prod_{n=1}^N(1-x^n)\prod_{m=1}^M(1-x^m)} \, .
\end{align}

\section{Cauchy's combinatorial identity}
\label{app:Cauchy}

We start with a combinatorial identity due to Cauchy  \cite[Corollary~2.6]{Andrews_1984}:
\begin{align}
    \frac{1}{(z;q)_\infty} = \sum_{k=0}^\infty \frac{q^{k^2-k} z^k}{(q;q)_k(z;q)_k} \, ,
\end{align}
and substitute $q=x$, $z=x^{N+1}$:
\begin{align}
    \frac{1}{(x^{N+1};x)_\infty} = \sum_{k=0}^\infty \frac{x^{(N+k)k}}{(x;x)_k(x^{N+1};x)_k} \, .
\end{align}
Using
\begin{align}
    (x;x)_N=\frac{(x;x)_\infty}{(x^{N+1};x)_\infty} \, ,
\end{align}
we get
\begin{align}
    \frac{1}{(x;x)_\infty} &= \sum_{k=0}^\infty \frac{x^{(N+k)k}}{(x;x)_k(x;x)_N(x^{N+1};x)_k} \\
    &= \sum_{k=0}^\infty \frac{x^{(N+k)k}}{(x;x)_k(x;x)_{N+k}} \, .
\end{align} 

\section{Koszul dual branes}
\label{app:branes}

Let us check in a few examples that $f^! \equiv -f/(1-f)$ is the single-letter index of a ``Koszul dual" brane in B-model, i.e. a probe brane transverse to the original stack in all directions. These branes correspond to twists of BPS branes in type IIB superstring theory, see e.g. \cite{Yoo:2025qlw}.

First, consider B-model branes in $\mathbb{C}^3$. The single-letter index of the worldvolume theory on a D$(-1)$-brane is
\begin{align}
    f_{D(-1)} &=1-(1-x)(1-y)(1-z) \, .
\end{align}
This is the index of the BMN truncation of $\mathcal{N}=4$ SYM. Then, %\footnote{GGE for the BMN index will be studied in \cite{inprogress2}.} 
\begin{align}
    f^!_{D(-1)} &= 1-\frac{1}{(1-x)(1-y)(1-z)} \, .
\end{align}
$f^!_{D(-1)}$ counts derivatives of a ghost with at least one derivative:
\begin{align}
    \partial^i_x\partial^j_y\partial^k_z c \, , \quad i+j+k>0 \, .
\end{align}
We identify it with the index of a D5-brane in $\mathbb{C}^3$:
\begin{align}
    f^!_{D(-1)} = f_{D5} \, .
\end{align}

Now consider D1-branes:
\begin{align}    
    f_{D1} &= 1-\frac{(1-x)(1-y)}{(1-q)} \, ,
\end{align}
with the one-loop BPS condition $q=xy$. This is the Schur index of $\mathcal{N}=4$ SYM. Then,
\begin{align}
    f^!_{D1} &= 1-\frac{(1-q)}{(1-x)(1-y)} \, .
\end{align}
This is the $1/4$ BPS index of $\mathcal{N}=1$ SYM with the one-loop BPS condition $q=xy$. We identify it with the index of a D3-brane in $\mathbb{C}^3$ (see e.g. \cite{Budzik:2023xbr}):
\begin{align}
    f^!_{D1} = f_{D3} \, .
\end{align}
Note that $f^{!!}=f$, so we are done with $\mathbb{C}^3$.

Now consider B-model branes in $\mathbb{C}^5$. The relevant branes for Murthy's GGE \cite{Murthy:2022ien} of $\mathcal{N}=4$ SYM index are D3- and D5-branes \cite{inprogress2}:
\begin{align}
    f_{D3} &= 1-\frac{(1-x)(1-y)(1-z)}{(1-p)(1-q)} \\
    f^!_{D3} &= 1-\frac{(1-p)(1-q)}{(1-x)(1-y)(1-z)} = f_{D5} \, ,
\end{align}
with the one-loop BPS condition $pq=xyz$. One can also check that $f^!_{D(-1)}=f_{D9}$ and $f^!_{D1}=f_{D7}$ in $\mathbb{C}^5$.

\end{appendix}

\bibliography{apssamp}% Produces the bibliography via BibTeX.

%In particular let us focus on coefficients of $t^{24}x^4y^4z^4$ The interesting case is $N=2$ since the BPS spectrum includes the so-called fortuitous operators \cite{Chang:2022mjp, Choi:2022caq, Choi:2023znd, Choi:2023vdm, Budzik:2023vtr,Chang:2023zqk}, which become BPS due to trace relations at special finite values of $N$. The first known $N=2$ fortuitous state arises at the order $\mathcal{O}(t^{24})$. The next correction $Z_{5|3}^+$ starts at $\mathcal{O}(t^{30})$.

\clearpage
\widetext
\renewcommand{\theequation}{S\arabic{equation}}

\section*{Supplemental Material}

In the supplemental material, we test the expansion \eqref{eq:inf} and the conjecture \eqref{eq:main} for the Hilbert series of superinvariants in multiple examples.

\section{Checks of the expansion}
\label{app:tables}

In this section, we verify the expansion \eqref{eq:inf} for the two matrix case $f=x+y$ (Table \ref{tab:2M}), the three matrix case $f=x+y+z$ (Table \ref{tab:3M}), the $1/4$ BPS index $f=x+y-xy$ (Table \ref{tab:14bps}) and the BMN index $f=1-(1-x)(1-y)(1-z)$ (Table \ref{tab:BMN}). 

We compute the integrals \eqref{eq:Zplus} by residues, keeping distinct fugacities but setting them equal for the tables. The computations start getting challenging around 5 or 6 integrals.

\begin{table}[H]
\centering
\begin{tabularx}{\textwidth}{|c | X|}
\hline
\multicolumn{1}{|c|}{$Z_{\infty}$} & $1 + 2 t + 6 t^2 + 14 t^3 + 34 t^4 + 74 t^5 + 166 t^6 + 350 t^7 + 746 t^8 + 1546 t^9 + 3206 t^{10} + 6550 t^{11} + 13386 t^{12} + 
 27114 t^{13} + 54894 t^{14} +  \dots$                                           \\ \hline
\multicolumn{2}{l}{$N=0$}                                                                                                                                           \\ \hline
\multicolumn{1}{|c|}{$Z_{1|1}$}    & $1 + 2 t + 6 t^2 + 14 t^3 + \color{blue} 28 t^4 + 50 t^5 + 82 t^6 + 126 t^7 + 184 t^8 + 258 t^9 + 350 t^{10} + 462 t^{11} + 596 t^{12}++754 t^{13} + 938 t^{14}+\dots$                                              \\ \hline
\multicolumn{1}{|c|}{$Z_{2|2}$}    & $1 + 2 t + 6 t^2 + 14 t^3 + 34 t^4 + 74 t^5 + 166 t^6 + 350 t^7 + 746 t^8 + \color{blue} 1526 t^9 + 3106 t^{10} + 6106 t^{11} + 11810 t^{12}++22114 t^{13} + 40518 t^{14}+\dots$                                            \\ \hline
\multicolumn{1}{|c|}{$Z_{3|3}$}    &  $1 + 2 t + 6 t^2 + 14 t^3 + 34 t^4 + 74 t^5 + 166 t^6 + 350 t^7 + 
 746 t^8 + 1546 t^9 + 3206 t^{10} + 6550 t^{11} + 13386 t^{12} + 
 27114 t^{13} + 54894 t^{14}+\dots$                                          \\ \hline
\multicolumn{2}{l}{$N=1$}                                                                                                                                           \\ \hline
\multicolumn{1}{|c|}{$Z_{1|0}$}    & $1 + 2 t + \color{blue} 3 t^2 + 4 t^3 + 5 t^4 + 6 t^5 + 7 t^6 + 8 t^7 + 9 t^8 + 
 10 t^9 + 11 t^{10} + 12 t^{11} + 13 t^{12} + 14 t^{13} + 15 t^{14}+\dots$                   \\ \hline
\multicolumn{1}{|c|}{$Z_{2|1}$}    & $1 + 2 t + 6 t^2 + 14 t^3 + 34 t^4 + 74 t^5 + \color{blue} 156 t^6 + 306 t^7 + 
 575 t^8 + 1024 t^9 + 1754 t^{10} + 2884 t^{11} + 4592 t^{12} + 7084 t^{13} + 10648 t^{14}+\dots$  \\ \hline
\multicolumn{1}{|c|}{$Z_{3|2}$}    & $1 + 2 t + 6 t^2 + 14 t^3 + 34 t^4 + 74 t^5 + 166 t^6 + 350 t^7 + 
 746 t^8 + 1546 t^9 + 3206 t^{10} + 6550 t^{11} + \color{blue} 13351 t^{12} + 26924 t^{13} + 53974 t^{14}+\dots$ \\ \hline
\multicolumn{2}{l}{$N=2$}                                                                                                                                           \\ \hline
\multicolumn{1}{|c|}{$Z_{2|0}$}    &  $1 + 2 t + 6 t^2 + \color{blue} 10 t^3 + 20 t^4 + 30 t^5 + 50 t^6 + 70 t^7 + 
 105 t^8 + 140 t^9 + 196 t^{10} + 252 t^{11} + 336 t^{12} + 420 t^{13} + 540 t^{14}+\dots$                                                                                                       \\ \hline
\multicolumn{1}{|c|}{$Z_{3|1}$}    &  $1 + 2 t + 6 t^2 + 14 t^3 + 34 t^4 + 74 t^5 + 166 t^6 + 350 t^7 + 
 \color{blue} 731 t^8 + 1476 t^9 + 2916 t^{10} + 5588 t^{11} + 10460 t^{12} + 
 19048 t^{13} + 33868 t^{14}+\dots$                                                                                                                                \\ \hline
\multicolumn{1}{|c|}{$Z_{4|2}$}    &  $1 + 2 t + 6 t^2 + 14 t^3 + 34 t^4 + 74 t^5 + 166 t^6 + 350 t^7 + 
 746 t^8 + 1546 t^9 + 3206 t^{10} + 6550 t^{11} + 13386 t^{12} + 
 27114 t^{13} + 54894 t^{14} +\dots$                                                                                                                                \\ \hline
\end{tabularx}
\caption{Two matrix case with $f=x+y$. We set $x=y=t$ in the table. For $N=0,1,2$ we check that $Z_{N+K|K}$ given by \eqref{eq:main} converges to $Z_\infty$ as $K$ increases.}
\label{tab:2M}
\end{table}

\begin{table}[H]
\centering
\begin{tabularx}{\textwidth}{|c | X|}
\hline
\multicolumn{1}{|c|}{$Z_{\infty}$} & $1 + 3 t + 12 t^2 + 39 t^3 + 129 t^4 + 399 t^5 + 1245 t^6 + 3783 t^7 +  11514 t^8 + 34734 t^9 + 104754 t^{10} + 314922 t^{11} + 946623 t^{12}+2842077 t^{13}+\dots$                                            \\ \hline
\multicolumn{2}{l}{$N=0$}                                                                                                                                           \\ \hline
\multicolumn{1}{|c|}{$Z_{1|1}$}    & $1 + 3 t + 12 t^2 + 39 t^3 + \color{blue}108 t^4 + 261 t^5 + 564 t^6 + 1113 t^7 + 2040 t^8 + 3519 t^9 + 5772 t^{10} + 9075 t^{11} + 13764 t^{12}+20241 t^{13}+\dots$                                             \\ \hline
\multicolumn{1}{|c|}{$Z_{2|2}$}    & $1 + 3 t + 12 t^2 + 39 t^3 + 129 t^4 + 399 t^5 + 1245 t^6 + 3783 t^7 + 11514 t^8 + \color{blue}34503 t^9 + 102246 t^{10} + 295863 t^{11} + 834063 t^{12}+2276013 t^{13}+\dots$                                           \\ \hline
%\multicolumn{1}{|c|}{$Z_{3|3}$}    &                                            \\ \hline
\multicolumn{2}{l}{$N=1$}                                                                                                                                           \\ \hline
\multicolumn{1}{|c|}{$Z_{1|0}$}    & $1 + 3 t + \color{blue}6 t^2 + 10 t^3 + 15 t^4 + 21 t^5 + 28 t^6 + 36 t^7 + 45 t^8 + 55 t^9 + 66 t^{10} + 78 t^{11} + 91 t^{12}+105 t^{13}+\dots$             \\ \hline
\multicolumn{1}{|c|}{$Z_{2|1}$}    & $1 + 3 t + 12 t^2 + 39 t^3 + 129 t^4 + 399 t^5 + \color{blue}1189 t^6 + 3330 t^7 + 8814 t^8 + 21972 t^9 + 51852 t^{10} + 116160 t^{11} + 248243 t^{12}+507993 t^{13}+\dots$ \\ \hline
\multicolumn{1}{|c|}{$Z_{3|2}$}    & $1 + 3 t + 12 t^2 + 39 t^3 + 129 t^4 + 399 t^5 + 1245 t^6 + 3783 t^7 + 
 11514 t^8 + 34734 t^9 + 104754 t^{10} + 314922 t^{11} + \color{blue} 945842 t^{12} + 2831484 t^{13}+\dots$ \\ \hline
\multicolumn{2}{l}{$N=2$}                                                                                                                                           \\ \hline
\multicolumn{1}{|c|}{$Z_{2|0}$}    &  $1 + 3 t + 12 t^2 + \color{blue}29 t^3 + 75 t^4 + 156 t^5 + 328 t^6 + 612 t^7 + 1134 t^8 + 1950 t^9 + 3312 t^{10} + 5346 t^{11} + 8514 t^{12}+13068 t^{13}+\dots$                                                                                                       \\ \hline
\multicolumn{1}{|c|}{$Z_{3|1}$}    &  $1 + 3 t + 12 t^2 + 39 t^3 + 129 t^4 + 399 t^5 + 1245 t^6 + 3783 t^7 +\color{blue} 11388 t^8 + 33551 t^9 + 96651 t^{10} + 270621 t^{11} + 735418 t^{12}+1935864 t^{13}+\dots$                                                                                                                             \\ \hline
%\multicolumn{1}{|c|}{$Z_{4|2}$}    &                                                                                                                              \\ \hline
\end{tabularx}
\caption{Three matrix case with $f=x+y+z$. We set $x=y=z=t$ in the table. For $N=0,1,2$ we check that $Z_{N+K|K}$ given by \eqref{eq:main} converges to $Z_\infty$ as $K$ increases.}
\label{tab:3M}
\end{table}

\FloatBarrier

\begin{table}[H]
\centering
\begin{tabularx}{\textwidth}{|c | X|}
\hline
\multicolumn{1}{|c|}{$Z_{\infty}$} & $1 + 2 t + 5 t^2 + 10 t^3 + 20 t^4 + 36 t^5 + 65 t^6 + 110 t^7 + 185 t^8 + 300 t^9 + 481 t^{10} + 752 t^{11} + 1165 t^{12}+1770 t^{13} + 2665 t^{14}+3956 t^{15}+\dots$                                            \\ \hline
\multicolumn{2}{l}{$N=0$}                                                                                                                                           \\ \hline
\multicolumn{1}{|c|}{$Z_{1|1}$}    & $1 + 2 t + 5 t^2 + 10 t^3 + \color{blue} 14 t^4 + 18 t^5 + 21 t^6 + 26 t^7 + 
 30 t^8 + 34 t^9 + 37 t^{10} + 42 t^{11} + 46 t^{12} + 50 t^{13} + 53 t^{14}+58 t^{15}+\dots$                                             \\ \hline
\multicolumn{1}{|c|}{$Z_{2|2}$}    & $1 + 2 t + 5 t^2 + 10 t^3 + 20 t^4 + 36 t^5 + 65 t^6 + 110 t^7 +  185 t^8 + \color{blue} 280 t^9 + 411 t^{10} + 556 t^{11} + 743 t^{12} + 968 t^{13} +  1261 t^{14}+1600 t^{15}+\dots$                                           \\ \hline
\multicolumn{1}{|c|}{$Z_{3|3}$}    & $1 + 2 t + 5 t^2 + 10 t^3 + 20 t^4 + 36 t^5 + 65 t^6 + 110 t^7 + 
 185 t^8 + 300 t^9 + 481 t^{10} + 752 t^{11} + 1165 t^{12} + 1770 t^{13} + 2665 t^{14} + 3956 t^{15} +\dots$                                           \\ \hline
\multicolumn{2}{l}{$N=1$}                                                                                                                                           \\ \hline
\multicolumn{1}{|c|}{$Z_{1|0}$}    & $1 + 2 t + \color{blue} 2 t^2 + 2 t^3 + 2 t^4 + 2 t^5 + 2 t^6 + 2 t^7 + 2 t^8 + 
 2 t^9 + 2 t^{10} + 2 t^{11} + 2 t^{12} + 2 t^{13} + 2 t^{14}+2t^{15}+\dots$             \\ \hline
\multicolumn{1}{|c|}{$Z_{2|1}$}    & $1 + 2 t + 5 t^2 + 10 t^3 + 20 t^4 + 36 t^5 + \color{blue} 55 t^6 + 78 t^7 + 
 102 t^8 + 132 t^9 + 168 t^{10} + 208 t^{11} + 252 t^{12} + 294 t^{13} + 
 343 t^{14}+400 t^{15}+\dots$ \\ \hline
\multicolumn{1}{|c|}{$Z_{3|2}$}    & $1 + 2 t + 5 t^2 + 10 t^3 + 20 t^4 + 36 t^5 + 65 t^6 + 110 t^7 + 
 185 t^8 + 300 t^9 + 481 t^{10} + 752 t^{11} + \color{blue} 1130 t^{12} + 1640 t^{13} + 
 2280 t^{14}+3078 t^{15}+\dots$ \\ \hline
\multicolumn{2}{l}{$N=2$}                                                                                                                                           \\ \hline
\multicolumn{1}{|c|}{$Z_{2|0}$}    &  $1 + 2 t + 5 t^2 + \color{blue} 6 t^3 + 9 t^4 + 10 t^5 + 13 t^6 + 14 t^7 + 17 t^8 + 
 18 t^9 + 21 t^{10} + 22 t^{11} + 25 t^{12} + 26 t^{13} + 29 t^{14}+30t^{15}+\dots$                                                                                                       \\ \hline
\multicolumn{1}{|c|}{$Z_{3|1}$}    &  $1 + 2 t + 5 t^2 + 10 t^3 + 20 t^4 + 36 t^5 + 65 t^6 + 110 t^7 + \color{blue} 170 t^8 + 250 t^9 + 346 t^{10} + 466 t^{11} + 613 t^{12} + 798 t^{13} + 1015 t^{14}+1268 t^{15}+\dots$                                                                                                                             \\ \hline
\multicolumn{1}{|c|}{$Z_{4|2}$}    &  $1 + 2 t + 5 t^2 + 10 t^3 + 20 t^4 + 36 t^5 + 65 t^6 + 110 t^7 + 
 185 t^8 + 300 t^9 + 481 t^{10} + 752 t^{11} + 1165 t^{12} + 1770 t^{13} + 2665 t^{14} + \color{blue} 3900 t^{15}+\dots$                                                                                                                             \\ \hline
\end{tabularx}
\caption{$1/4$ BPS case with $f=x+y-xy$. We set $x=y=t$ in the table. For $N=0,1,2$ we check that $Z_{N+K|K}$ given by \eqref{eq:main} converges to $Z_\infty$ as $K$ increases.}
\label{tab:14bps}
\end{table} 

\vfill

\begin{table}[H]
\centering
\begin{tabularx}{\textwidth}{|c | X|}
\hline
\multicolumn{1}{|c|}{$Z_{\infty}$} & $1+3 t^2+9 t^4+22 t^6+51 t^8+108 t^{10}+221 t^{12}+429 t^{14}+810 t^{16}+1479 t^{18}+2640 t^{20}+4599 t^{22}+7868 t^{24}+\dots$                                            \\ \hline
\multicolumn{2}{l}{$N=0$}                                                                                                                                           \\ \hline
\multicolumn{1}{|c|}{$Z_{1|1}$}    &  $1 + 3 t^2 + 9 t^4 + 22 t^6 + {\color{blue} 30 t^8 + 27 t^{10} + 25 t^{12} + 57 t^{14} + 
 90 t^{16} + 48 t^{18} - 15 t^{20} + 93 t^{22} + 248 t^{24}+\dots}$                                           \\ \hline
\multicolumn{1}{|c|}{$Z_{2|2}$}    &  $1 + 3 t^2 + 9 t^4 + 22 t^6 + 51 t^8 + 108 t^{10} + 221 t^{12} + 
 429 t^{14} + 810 t^{16} + \color{blue} 1248 t^{18} + 1419 t^{20} + 1017 t^{22} + 1090 t^{24}+\dots$                                          \\ \hline
\multicolumn{2}{l}{$N=1$}                                                                                                                                           \\ \hline
\multicolumn{1}{|c|}{$Z_{1|0}$}    &  $1 + 3 t^2 + \color{blue} 3 t^4 + 2 t^6 + 3 t^8 + 3 t^{10} + 2 t^{12} + 3 t^{14} + 
 3 t^{16} + 2 t^{18} + 3 t^{20} + 3 t^{22} + 2 t^{24}+\dots$            \\ \hline
\multicolumn{1}{|c|}{$Z_{2|1}$}    & $1 + 3 t^2 + 9 t^4 + 22 t^6 + 51 t^8 + 108 t^{10} + \color{blue} 165 t^{12} + 
 183 t^{14} + 162 t^{16} + 248 t^{18} + 594 t^{20} + 930 t^{22} + 587 t^{24}+\dots$ \\ \hline
\multicolumn{2}{l}{$N=2$}                                                                                                                                           \\ \hline
\multicolumn{1}{|c|}{$Z_{2|0}$}    &  $1 + 3 t^2 + 9 t^4 + \color{blue} 12 t^6 + 15 t^8 + 15 t^{10} + 24 t^{12} + 30 t^{14} + 
 27 t^{16} + 25 t^{18} + 42 t^{20} + 48 t^{22} + 36 t^{24}+\dots $           \\ \hline
\multicolumn{1}{|c|}{$Z_{3|1}$}    & $1 + 3 t^2 + 9 t^4 + 22 t^6 + 51 t^8 + 108 t^{10} + 221 t^{12} + 
 429 t^{14} + \color{blue} 684 t^{16} + 863 t^{18} + 849 t^{20} + 945 t^{22} + 1921 t^{24}+\dots $  \\ \hline
\end{tabularx}
\caption{BMN truncation of $\mathcal{N}=4$ SYM with $f=1-(1-x)(1-y)(1-z)$. We set $x=y=z=t^2$ in the table. For $N=0,1,2$ we check that $Z_{N+K|K}$ given by \eqref{eq:main} converges to $Z_\infty$ as $K$ increases.}
\label{tab:BMN}
\end{table}

\vfill

An important check is the BMN index for $N=2$ since the BPS spectrum includes fortuitous operators \cite{Chang:2022mjp, Choi:2022caq, Choi:2023znd, Choi:2023vdm, Budzik:2023vtr,Chang:2023zqk}, which become BPS due to trace relations at special finite values of $N$. We were not able to compute the full $Z_{4|2}^+$ integral (it involves $\sim 300 \, 000$ residues) but we compute the series coefficient of $x^4y^4z^4\sim t^{24}$, the order at which the first known $N=2$ fortuitous state arises, see Table \ref{tab:t24}.

\vfill

\begin{table}[H]
\centering
\begin{tabular}{|c|ccc|}
\hline
$Z_\infty$ & $Z_{2|0}$ & $Z_{3|1}^+$ & $Z_{4|2}^+$ \\ \hline
125        & -3        & 46          &  82           \\ \hline
\end{tabular}
\caption{Coefficients of $x^4y^4z^4\sim t^{24}$. The next correction $Z_{5|3}^+$ starts at $\mathcal{O}(t^{30})$.}
\label{tab:t24}
\end{table}

\FloatBarrier

\clearpage
\section{Two matrices}
\label{app:2}

In this section, we present explicit results for the two matrix case, $f=x+y$. We compute the integrals \eqref{eq:Zplus} for small values of $N$ and $M$, shown in \eqref{eq:2Mfirst}-\eqref{eq:2Mlast}. From these, the conjectured Hilbert series of superinvariants are obtained via the sums \eqref{eq:main}.

The integrals \eqref{eq:Zplus} for the two matrix case, $f=x+y$, are:
\begin{align}
    Z_{1|1}^+ =& \frac{P_{1|1}^+}{(1-x)^2 (1- y)^2} \label{eq:2Mfirst} \\
    Z_{2|2}^+ =& \frac{P_{2|2}^+}{(1-x)^2 (1-x^2)^2 (1-y)^2 (1-y^2)^2 (1-x y)^2} \\
    Z_{1} =& \frac{1}{(1-x)(1-y)} \\
    Z_{2|1}^+ =& \frac{P_{2|1}^+}{(1-x)^2 (1-x^2) (1-y)^2 (1-y^2) (1-xy)} \\
    Z_{2} =& \frac{1}{(1 - x) (1 - x^2) (1 - y) (1 - y^2) (1 - x y)} \\
    Z_{3|1}^+ =& \frac{P_{3|1}^+}{(1 - x)^2 (1 - x^2) (1 -x^3) (1 - y)^2 (1 - y^2)(1-y^3) (1 - x y)^2 (1 - x^2 y) (1 - x y^2)}  \, , \label{eq:2Mlast}
\end{align}
with the numerators:
\begin{align}
P_{1|1}^+ &= x + y - 2 x y + x^2 y + x y^2 \\
P_{2|2}^+ &= x^4 + x^3 y + 2 x^2 y^2 + x y^3 + y^4 -2 x^5 y - 3 x^4 y^2 - 2 x^3 y^3 - 3 x^2 y^4 - 2 x y^5 + 2 x^5 y^2 + 2 x^4 y^3 + 2 x^3 y^4 \nonumber \\
&+ 2 x^2 y^5 + x^7 y + 3 x^6 y^2 + 5 x^5 y^3 + 8 x^4 y^4 + 5 x^3 y^5 + 
    3 x^2 y^6 + x y^7 + 2 x^6 y^3 + 2 x^3 y^6 + 
 4 x^5 y^5 + 2 x^7 y^4 \nonumber \\
 &+ 4 x^6 y^5 + 4 x^5 y^6 + 2 x^4 y^7 + 2 x^8 y^4 + x^7 y^5 + 4 x^6 y^6 + x^5 y^7 + 2 x^4 y^8 + 2 x^8 y^5 + 2 x^7 y^6 + 2 x^6 y^7 + 2 x^5 y^8 \nonumber   \\
 &+ x^8 y^6 + 2 x^7 y^7 + x^6 y^8 \\
P_{2|1}^+ &= x^2 + x y + y^2 - x^2 y - x y^2 - x^3 y - x y^3 +x^4 y + 2 x^3 y^2 + 2 x^2 y^3 + x y^4 + 2 x^3 y^3 - x^4 y^3 - x^3 y^4 + x^4 y^4 \\
P_{3|1}^+ &= x^3 + x^2 y + x y^2 + y^3 -x^3 y - x y^3 -x^4 y - x^3 y^2 - x^2 y^3 - x y^4 -x^5 y + x^4 y^2 + x^2 y^4 - x y^5 + x^6 y + 2 x^5 y^2 \nonumber \\
&+ 3 x^4 y^3 + 3 x^3 y^4 + 2 x^2 y^5 + x y^6 +2 x^6 y^2 + 2 x^5 y^3 + 4 x^4 y^4 + 2 x^3 y^5 + 2 x^2 y^6 -x^7 y^2 - x^5 y^4 - x^4 y^5 - x^2 y^7 \nonumber \\
&-x^7 y^3 - x^6 y^4 - x^4 y^6 - x^3 y^7 -x^7 y^4 - x^6 y^5 - x^5 y^6 - x^4 y^7 + 2 x^8 y^4 + 3 x^7 y^5 + 2 x^6 y^6 + 3 x^5 y^7 + 2 x^4 y^8 \nonumber \\
&-x^8 y^5 + x^7 y^6 + x^6 y^7 - x^5 y^8 -2 x^8 y^7 - 2 x^7 y^8 + 2 x^8 y^8   \, .     \end{align}

We verify these computations by explicitly enumerating all polynomials in (super)traces of two matrices $X,Y$ up to length six using Mathematica. The resulting Hilbert series expansions are collected in Table~\ref{tab:2Mops}.
 
\begin{table}[H]
\centering
\begin{tabularx}{\textwidth}{|c | X|}
\multicolumn{2}{l}{$N=0$}  \\ \hline
\multicolumn{1}{|c|}{$U(1|1)$}    & $1 + x + y + 2 x^2 + 2 x y + 2 y^2 + 3 x^3 + 4 x^2 y + 4 x y^2 + 3 y^3 + 4 x^4 + 6 x^3 y + 8 x^2 y^2 + 6 x y^3 + 4 y^4 + 5 x^5 + 8 x^4 y + 12 x^3 y^2 + 12 x^2 y^3 + 8 x y^4 + 5 y^5 + 6 x^6 + 10 x^5 y + 16 x^4 y^2 + 18 x^3 y^3 + 16 x^2 y^4 + 10 x y^5 + 6 y^6+\dots$                                              \\ \hline
\multicolumn{1}{|c|}{$U(2|2)$}    & $1 + x + y + 2 x^2 + 2 x y + 2 y^2 + 3 x^3 + 4 x^2 y + 4 x y^2 + 3 y^3 + 5 x^4 + 7 x^3 y + 10 x^2 y^2 + 7 x y^3 + 5 y^4 + 7 x^5 + 12 x^4 y + 18 x^3 y^2 + 18 x^2 y^3 + 12 x y^4 + 7 y^5 +11 x^6 + 19 x^5 y + 34 x^4 y^2 + 38 x^3 y^3 + 34 x^2 y^4 + 
    19 x y^5 + 11 y^6+\dots$                                            \\ \hline
\multicolumn{2}{l}{$N=1$}   \\ \hline
\multicolumn{1}{|c|}{$U(1)$}    & $1 + x + y + x^2 + x y + y^2 + x^3 + x^2 y + x y^2 + y^3 + x^4 + x^3 y + x^2 y^2 + x y^3 + y^4 + x^5 + x^4 y + x^3 y^2 + x^2 y^3 + x y^4 + y^5 + x^6 + x^5 y + x^4 y^2 + x^3 y^3 + x^2 y^4 + x y^5 + y^6+\dots$                   \\ \hline
\multicolumn{1}{|c|}{$U(2|1)$}    & $1 + x + y + 2 x^2 + 2 x y + 2 y^2 + 3 x^3 + 4 x^2 y + 4 x y^2 + 3 y^3 + 5 x^4 + 7 x^3 y + 10 x^2 y^2 + 7 x y^3 + 5 y^4 + 7 x^5 + 12 x^4 y + 18 x^3 y^2 + 18 x^2 y^3 + 12 x y^4 + 7 y^5 +10 x^6 + 18 x^5 y + 32 x^4 y^2 + 36 x^3 y^3 + 32 x^2 y^4 + 
    18 x y^5 + 10 y^6+\dots$  \\ \hline
\end{tabularx}
\end{table}

\begin{table}[H]
\centering
\begin{tabularx}{\textwidth}{|c | X|}
\multicolumn{2}{l}{$N=2$}    \\ \hline
\multicolumn{1}{|c|}{$U(2)$}    &  $1 + x + y + 2 x^2 + 2 x y + 2 y^2 + 2 x^3 + 3 x^2 y + 3 x y^2 + 2 y^3 + 3 x^4 + 4 x^3 y + 6 x^2 y^2 + 4 x y^3 + 3 y^4 + 3 x^5 + 5 x^4 y + 7 x^3 y^2 + 7 x^2 y^3 + 5 x y^4 + 3 y^5 +4 x^6 + 6 x^5 y + 10 x^4 y^2 + 10 x^3 y^3 + 10 x^2 y^4 + 
    6 x y^5 + 4 y^6+\dots$                                                                                                       \\ \hline
\multicolumn{1}{|c|}{$U(3|1)$}    &  $1 + x + y + 2 x^2 + 2 x y + 2 y^2 + 3 x^3 + 4 x^2 y + 4 x y^2 + 3 y^3 + 5 x^4 + 7 x^3 y + 10 x^2 y^2 + 7 x y^3 + 5 y^4 + 7 x^5 + 12 x^4 y + 18 x^3 y^2 + 18 x^2 y^3 + 12 x y^4 + 7 y^5 +11 x^6 + 19 x^5 y + 34 x^4 y^2 + 38 x^3 y^3 + 34 x^2 y^4 + 
    19 x y^5 + 11 y^6+\dots$                                                                                                                                \\ \hline
\end{tabularx}
\caption{Hilbert series counting polynomials in (super)traces in two matrices $X$ and $Y$, obtained by explicitly enumerating polynomials in (super)traces up to length six in Mathematica.}
\label{tab:2Mops}
\end{table}

\section{Three matrices}
\label{app:3}

\setcounter{equation}{10}

In this section, we present explicit results for the three matrix case, $f=x+y+z$. We compute the integrals \eqref{eq:Zplus} for small values of $N$ and $M$, shown in \eqref{eq:3Mfirst}-\eqref{eq:3Mlast}. From these, the conjectured Hilbert series of superinvariants are obtained via the sums \eqref{eq:main}.

The integrals \eqref{eq:Zplus} for the three matrix case, $f=x+y+z$, are:

\begin{align}
    Z_{1|1}^+ =& \frac{P_{1|1}^+}{(1- x)^2 (1- y)^2 (1- z)^2} \label{eq:3Mfirst} \\
    Z_{2|2}^+ =& \frac{P_{2|2}^+}{(1- x)^2 (1 - x^2)^2 (1- y)^2 (1 - y^2)^2 (1 - z)^2 (1 - z^2)^2 (1- x y)^2  (1 - x z)^2 (1 - y z)^2} \\
    Z_{1} =& \frac{1}{(1-x)(1-y)(1-z)} \\
    Z_{2|1}^+ =& \frac{P_{2|1}^+}{(1 - x)^2 (1 - x^2) (1 - y)^2 (1 - y^2)  (1 - z)^2 (1 - z^2) (1 - x y) (1 - x z) (1 - y z) } \\
     Z_2 =& \frac{1 + x y z}{(1 - x) (1 - x^2) (1 - y) (1 - y^2) (1- z)(1 - z^2) (1 - x y)  (1 - x z) (1 - y z)} \\
    Z_{3|1}^+ =& \frac{P_{3|1}^+}{(1- x)^2 (1 - x^2) (1-x^3) (1- y)^2 (1 - y^2)(1-y^3)(1- z)^2 (1 - z^2) (1-z^3) (1- x y)^2} \\
    &\times \frac{1}{(1- x^2 y) (1- x y^2) (1- x z)^2 (1- x^2 z)(1- x z^2)(1- y z)^2 (1- y^2 z)  (1- y z^2) (1 - x y z)} \label{eq:3Mlast} \, ,
\end{align}
with the numerators:\footnote{{\setlength{\fboxrule}{0pt}\fbox{$P_{2|2}^+, P_{2|1}^+, P_{3|1}^+$ have $999,158$ and $1922$ terms, respectively. We show terms up to $\mathcal{O}(t^6)$ needed for comparison with Table \ref{tab:3Mops}.}}}

\begin{align}
P_{1|1}^+ &= x + y + z-2 x y - 2 x z - 2 y z +x^2 y + x y^2 + x^2 z + 6 x y z + y^2 z + x z^2 + y z^2 -2 x^2 y z - 2 x y^2 z - 2 x y z^2 \nonumber \\
&+ x^2 y^2 z + x^2 y z^2 + x y^2 z^2 \\
P_{2|2}^+ &= x^4 + x^3 y + 2 x^2 y^2 + x y^3 + y^4 + x^3 z + 2 x^2 y z + 
 2 x y^2 z + y^3 z + 2 x^2 z^2 + 2 x y z^2 + 2 y^2 z^2 + x z^3 + 
 y z^3 + z^4 \nonumber \\
 &+ 2 x^3 y z + 2 x^2 y^2 z + 2 x y^3 z + 2 x^2 y z^2 + 
 2 x y^2 z^2 + 2 x y z^3 - 2 x^5 y - 3 x^4 y^2 - 2 x^3 y^3 - 
 3 x^2 y^4 - 2 x y^5 - 2 x^5 z \nonumber \\
 &- 4 x^4 y z - 5 x^3 y^2 z - 
 5 x^2 y^3 z - 4 x y^4 z - 2 y^5 z - 3 x^4 z^2 - 5 x^3 y z^2 - 
 6 x^2 y^2 z^2 - 5 x y^3 z^2 - 3 y^4 z^2 - 2 x^3 z^3 \nonumber \\
 &- 5 x^2 y z^3 - 
 5 x y^2 z^3 - 2 y^3 z^3 - 3 x^2 z^4 - 4 x y z^4 - 3 y^2 z^4 - 
 2 x z^5 - 2 y z^5+\dots
\end{align}
\begin{align}
P_{2|1}^+ &= x^2 + x y + y^2 + x z + y z + z^2-x^2 y - x y^2 - x^2 z - 
    x y z - y^2 z - x z^2 - y z^2-x^3 y - x y^3 - x^3 z - 
    y^3 z - x z^3 \nonumber \\
    &- y z^3 + x^4 y + 2 x^3 y^2 + 2 x^2 y^3 + 
    x y^4 + x^4 z + 5 x^3 y z + 6 x^2 y^2 z + 5 x y^3 z + y^4 z + 
    2 x^3 z^2 + 6 x^2 y z^2 + 6 x y^2 z^2 \nonumber \\
    &+ 2 y^3 z^2 + 2 x^2 z^3 + 
    5 x y z^3 + 2 y^2 z^3 + x z^4 + y z^4 + 2 x^3 y^3 + 
    x^4 y z + 2 x^3 y^2 z + 2 x^2 y^3 z + x y^4 z + 2 x^3 y z^2 + 
    2 x^2 y^2 z^2 \nonumber \\ 
    &+ 2 x y^3 z^2+ 2 x^3 z^3 + 2 x^2 y z^3 + 
    2 x y^2 z^3 + 2 y^3 z^3 + x y z^4+\dots \\ 
P_{3|1}^+ &= x^3 + x^2 y + x y^2 + y^3 + x^2 z + x y z + y^2 z + x z^2 +     y z^2 + z^3 -x^3 y - x y^3 - x^3 z - y^3 z - x z^3 - y z^3 -x^4 y \nonumber \\
&- x^3 y^2 - x^2 y^3 - x y^4 - x^4 z - x^3 y z - 2 x^2 y^2 z - x y^3 z - y^4 z - x^3 z^2 - 2 x^2 y z^2 -    2 x y^2 z^2 - y^3 z^2 - x^2 z^3 - x y z^3 \nonumber \\
&- y^2 z^3 - x z^4 -    y z^4-x^5 y + x^4 y^2 + x^2 y^4 - x y^5 - x^5 z + 2 x^4 y z +     2 x^3 y^2 z + 2 x^2 y^3 z + 2 x y^4 z - y^5 z + x^4 z^2 \nonumber \\
&+    2 x^3 y z^2 + 5 x^2 y^2 z^2 + 2 x y^3 z^2 + y^4 z^2 +    2 x^2 y z^3 + 2 x y^2 z^3 + x^2 z^4 + 2 x y z^4 + y^2 z^4 -    x z^5 - y z^5+\dots
\end{align}

We verify these computations by explicitly enumerating all polynomials in (super)traces of three matrices $X,Y,Z$ up to length six using Mathematica. The resulting Hilbert series expansions are collected in Table~\ref{tab:3Mops}.

\begin{table}[H]
\centering
\begin{tabularx}{\textwidth}{|c | X|}
\multicolumn{2}{l}{$N=0$}  \\ \hline
\multicolumn{1}{|c|}{$U(1|1)$}    & $1 +~x +~y +~z +~2 x^2 +~2 x y +~2 y^2 +~2 x z +~2 y z +~2 z^2 + 
 3 x^3 + 4 x^2 y + 4 x y^2 + 3 y^3 + 4 x^2 z + 6 x y z + 
    4 y^2 z + 4 x z^2 + 4 y z^2 + 3 z^3 + 
 4 x^4 + 6 x^3 y + 8 x^2 y^2 + 6 x y^3 + 4 y^4 + 6 x^3 z +~12 x^2 y z +~12 x y^2 z +~6 y^3 z +~8 x^2 z^2 +~12 x y z^2 +~8 y^2 z^2 + 6 x z^3 + 6 y z^3 + 4 z^4 + 5 x^5 + 8 x^4 y +~12 x^3 y^2 +~12 x^2 y^3 + 8 x y^4 + 5 y^5 + 
    8 x^4 z + 18 x^3 y z + 24 x^2 y^2 z + 18 x y^3 z + 8 y^4 z + 
    12 x^3 z^2 + 24 x^2 y z^2 + 24 x y^2 z^2 + 12 y^3 z^2 + 
    12 x^2 z^3 + 18 x y z^3 + 12 y^2 z^3 + 8 x z^4 + 8 y z^4 + 
    5 z^5 + 6 x^6 + 10 x^5 y + 16 x^4 y^2 + 18 x^3 y^3 + 16 x^2 y^4 + 
    10 x y^5 + 6 y^6 + 10 x^5 z + 24 x^4 y z + 36 x^3 y^2 z + 
    36 x^2 y^3 z + 24 x y^4 z + 10 y^5 z + 16 x^4 z^2 + 
    36 x^3 y z^2 + 48 x^2 y^2 z^2 + 36 x y^3 z^2 + 16 y^4 z^2 + 
    18 x^3 z^3 + 36 x^2 y z^3 + 36 x y^2 z^3 + 18 y^3 z^3 + 
    16 x^2 z^4 + 24 x y z^4 + 16 y^2 z^4 + 10 x z^5 + 10 y z^5 + 
    6 z^6+\dots$                                              \\ \hline
\multicolumn{1}{|c|}{$U(2|2)$}    & $1 + x + y + z + 
 2 x^2 + 2 x y + 2 y^2 + 2 x z + 2 y z + 2 z^2 + 
 3 x^3 + 4 x^2 y + 4 x y^2 + 3 y^3 + 4 x^2 z + 6 x y z + 
    4 y^2 z + 4 x z^2 + 4 y z^2 + 3 z^3 + 
 5 x^4 + 7 x^3 y + 10 x^2 y^2 + 7 x y^3 + 5 y^4 + 7 x^3 z + 
    14 x^2 y z + 14 x y^2 z + 7 y^3 z + 10 x^2 z^2 + 14 x y z^2 + 
    10 y^2 z^2 + 7 x z^3 + 7 y z^3 + 5 z^4 + 
 7 x^5 + 12 x^4 y + 18 x^3 y^2 + 18 x^2 y^3 + 12 x y^4 + 7 y^5 + 
    12 x^4 z + 28 x^3 y z + 38 x^2 y^2 z + 28 x y^3 z + 12 y^4 z + 
    18 x^3 z^2 + 38 x^2 y z^2 + 38 x y^2 z^2 + 18 y^3 z^2 + 
    18 x^2 z^3 + 28 x y z^3 + 18 y^2 z^3 + 12 x z^4 + 12 y z^4 + 
    7 z^5 + 
11 x^6 + 19 x^5 y + 34 x^4 y^2 + 38 x^3 y^3 + 34 x^2 y^4 + 
    19 x y^5 + 11 y^6 + 19 x^5 z + 52 x^4 y z + 84 x^3 y^2 z + 
    84 x^2 y^3 z + 52 x y^4 z + 19 y^5 z + 34 x^4 z^2 + 
    84 x^3 y z^2 + 120 x^2 y^2 z^2 + 84 x y^3 z^2 + 34 y^4 z^2 + 
    38 x^3 z^3 + 84 x^2 y z^3 + 84 x y^2 z^3 + 38 y^3 z^3 + 
    34 x^2 z^4 + 52 x y z^4 + 34 y^2 z^4 + 19 x z^5 + 19 y z^5 + 
    11 z^6+\dots$                                            \\ \hline
\multicolumn{2}{l}{$N=1$}   \\ \hline
\multicolumn{1}{|c|}{$U(1)$}    & $1 + x + y + z + x^2 + x y + y^2 + x z + y z + z^2 + 
 x^3 + x^2 y + x y^2 + y^3 + x^2 z + x y z + y^2 z + x z^2 + 
    y z^2 + z^3 + 
x^4 + x^3 y + x^2 y^2 + x y^3 + y^4 + x^3 z + x^2 y z + 
    x y^2 z + y^3 z + x^2 z^2 + x y z^2 + y^2 z^2 + x z^3 + y z^3 + 
    z^4 +x^5 + x^4 y + x^3 y^2 + x^2 y^3 + x y^4 + y^5 + 
    x^4 z + x^3 y z + x^2 y^2 z + x y^3 z + y^4 z + x^3 z^2 + 
    x^2 y z^2 + x y^2 z^2 + y^3 z^2 + x^2 z^3 + x y z^3 + y^2 z^3 + 
    x z^4 + y z^4 + z^5 + 
 x^6 + x^5 y + x^4 y^2 + x^3 y^3 + x^2 y^4 + x y^5 + y^6 + 
    x^5 z + x^4 y z + x^3 y^2 z + x^2 y^3 z + x y^4 z + y^5 z + 
    x^4 z^2 + x^3 y z^2 + x^2 y^2 z^2 + x y^3 z^2 + y^4 z^2 + 
    x^3 z^3 + x^2 y z^3 + x y^2 z^3 + y^3 z^3 + x^2 z^4 + x y z^4 + 
    y^2 z^4 + x z^5 + y z^5 + z^6+\dots$                   \\ \hline
\multicolumn{1}{|c|}{$U(2|1)$}    & $1 + x + y + z + 
 2 x^2 + 2 x y + 2 y^2 + 2 x z + 2 y z + 2 z^2 + 
 3 x^3 + 4 x^2 y + 4 x y^2 + 3 y^3 + 4 x^2 z + 6 x y z + 
    4 y^2 z + 4 x z^2 + 4 y z^2 + 3 z^3 + 
 5 x^4 + 7 x^3 y + 10 x^2 y^2 + 7 x y^3 + 5 y^4 + 7 x^3 z + 
    14 x^2 y z + 14 x y^2 z + 7 y^3 z + 10 x^2 z^2 + 14 x y z^2 + 
    10 y^2 z^2 + 7 x z^3 + 7 y z^3 + 5 z^4 + 
 7 x^5 + 12 x^4 y + 18 x^3 y^2 + 18 x^2 y^3 + 12 x y^4 + 7 y^5 + 
    12 x^4 z + 28 x^3 y z + 38 x^2 y^2 z + 28 x y^3 z + 12 y^4 z + 
    18 x^3 z^2 + 38 x^2 y z^2 + 38 x y^2 z^2 + 18 y^3 z^2 + 
    18 x^2 z^3 + 28 x y z^3 + 18 y^2 z^3 + 12 x z^4 + 12 y z^4 + 
    7 z^5 + 
10 x^6 + 18 x^5 y + 32 x^4 y^2 + 36 x^3 y^3 + 32 x^2 y^4 + 
    18 x y^5 + 10 y^6 + 18 x^5 z + 50 x^4 y z + 81 x^3 y^2 z + 
    81 x^2 y^3 z + 50 x y^4 z + 18 y^5 z + 32 x^4 z^2 + 
    81 x^3 y z^2 + 115 x^2 y^2 z^2 + 81 x y^3 z^2 + 32 y^4 z^2 + 
    36 x^3 z^3 + 81 x^2 y z^3 + 81 x y^2 z^3 + 36 y^3 z^3 + 
    32 x^2 z^4 + 50 x y z^4 + 32 y^2 z^4 + 18 x z^5 + 18 y z^5 + 
    10 z^6+\dots$  \\ \hline
\multicolumn{2}{l}{$N=2$}    \\ \hline
\multicolumn{1}{|c|}{$U(2)$}    &  $1 + x + y + z + 
 2 x^2 + 2 x y + 2 y^2 + 2 x z + 2 y z + 2 z^2 + 
 2 x^3 + 3 x^2 y + 3 x y^2 + 2 y^3 + 3 x^2 z + 5 x y z + 
    3 y^2 z + 3 x z^2 + 3 y z^2 + 2 z^3 + 
 3 x^4 + 4 x^3 y + 6 x^2 y^2 + 4 x y^3 + 3 y^4 + 4 x^3 z + 
    8 x^2 y z + 8 x y^2 z + 4 y^3 z + 6 x^2 z^2 + 8 x y z^2 + 
    6 y^2 z^2 + 4 x z^3 + 4 y z^3 + 3 z^4 + 
 3 x^5 + 5 x^4 y + 7 x^3 y^2 + 7 x^2 y^3 + 5 x y^4 + 3 y^5 + 
    5 x^4 z + 11 x^3 y z + 14 x^2 y^2 z + 11 x y^3 z + 5 y^4 z + 
    7 x^3 z^2 + 14 x^2 y z^2 + 14 x y^2 z^2 + 7 y^3 z^2 + 7 x^2 z^3 + 
    11 x y z^3 + 7 y^2 z^3 + 5 x z^4 + 5 y z^4 + 3 z^5 + 
 4 x^6 + 6 x^5 y + 10 x^4 y^2 + 10 x^3 y^3 + 10 x^2 y^4 + 
    6 x y^5 + 4 y^6 + 6 x^5 z + 14 x^4 y z + 20 x^3 y^2 z + 
    20 x^2 y^3 z + 14 x y^4 z + 6 y^5 z + 10 x^4 z^2 + 20 x^3 y z^2 + 
    28 x^2 y^2 z^2 + 20 x y^3 z^2 + 10 y^4 z^2 + 10 x^3 z^3 + 
    20 x^2 y z^3 + 20 x y^2 z^3 + 10 y^3 z^3 + 10 x^2 z^4 + 
    14 x y z^4 + 10 y^2 z^4 + 6 x z^5 + 6 y z^5 + 4 z^6+\dots$                                                                                                       \\ \hline
\multicolumn{1}{|c|}{$U(3|1)$}    &  $1 + x + y + z + 
 2 x^2 + 2 x y + 2 y^2 + 2 x z + 2 y z + 2 z^2 + 
 3 x^3 + 4 x^2 y + 4 x y^2 + 3 y^3 + 4 x^2 z + 6 x y z + 
    4 y^2 z + 4 x z^2 + 4 y z^2 + 3 z^3 + 
 5 x^4 + 7 x^3 y + 10 x^2 y^2 + 7 x y^3 + 5 y^4 + 7 x^3 z + 
    14 x^2 y z + 14 x y^2 z + 7 y^3 z + 10 x^2 z^2 + 14 x y z^2 + 
    10 y^2 z^2 + 7 x z^3 + 7 y z^3 + 5 z^4 + 
 7 x^5 + 12 x^4 y + 18 x^3 y^2 + 18 x^2 y^3 + 12 x y^4 + 7 y^5 + 
    12 x^4 z + 28 x^3 y z + 38 x^2 y^2 z + 28 x y^3 z + 12 y^4 z + 
    18 x^3 z^2 + 38 x^2 y z^2 + 38 x y^2 z^2 + 18 y^3 z^2 + 
    18 x^2 z^3 + 28 x y z^3 + 18 y^2 z^3 + 12 x z^4 + 12 y z^4 + 
    7 z^5 + 
 11 x^6 + 19 x^5 y + 34 x^4 y^2 + 38 x^3 y^3 + 34 x^2 y^4 + 
    19 x y^5 + 11 y^6 + 19 x^5 z + 52 x^4 y z + 84 x^3 y^2 z + 
    84 x^2 y^3 z + 52 x y^4 z + 19 y^5 z + 34 x^4 z^2 + 
    84 x^3 y z^2 + 120 x^2 y^2 z^2 + 84 x y^3 z^2 + 34 y^4 z^2 + 
    38 x^3 z^3 + 84 x^2 y z^3 + 84 x y^2 z^3 + 38 y^3 z^3 + 
    34 x^2 z^4 + 52 x y z^4 + 34 y^2 z^4 + 19 x z^5 + 19 y z^5 + 
    11 z^6+\dots$                                                                                                                                \\ \hline
\end{tabularx}
\caption{Hilbert series counting polynomials in (super)traces in three matrices $X$, $Y$ and $Z$, obtained by explicitly enumerating polynomials in (super)traces up to length six in Mathematica.}
\label{tab:3Mops}
\end{table}

\FloatBarrier

\section{Vectors}
\label{app:vectors}

\setcounter{equation}{21}

In this section, we compute the integrals \eqref{eq:main} for $(\bar{F},F)$ vectors \eqref{eq:hbarh} and verify the truncated expansion \eqref{eq:FbarF}.

The integrals $Z_{N|M}^+$ are much simpler for vectors than for adjoints so higher checks are possible but we do not include them to avoid repetition.

\subsection{$(\bar{F},F)=(1,1)$}

We perform the integrals \eqref{eq:ZPlusvector} by residues. For each $u_i$, the only pole inside the contour is $u_i^*=0$. After performing $N+1$ integrals over $u_1,\dots,u_{N+1}$, we find that $u_{N+2}$ has no pole inside the contour. Therefore,
\begin{align}
    Z_{N+k|k}^+ = 0 \, , \quad k\geq 2 \, .
\end{align}
For $k=1$, we can flip the contour for $v_1$ leaving only one pole $v_1^*=1/s$, which gives:
\begin{align}
    Z_{N+1|1}^+ &= \frac{(\bar{s}s)^{N+1}}{1-\bar{s}s} \, .
\end{align}
We have tested the above statements and the final formula for $N=0,1,2,3$ and $k=1,2,3$.

%\subsection{$(\bar{F},F)=(1,2)$}

%The $U(N)$ generating functions are
%\begin{align}
%    Z_0 &= 1 \\
%    Z_1 &= 1 + \bar{s} s_1 + \bar{s} s_2 \\
%    Z_2 &= 1 + \bar{s} s_1 + \bar{s} s_2 + \bar{s}^2 s_1^2 + \bar{s}^2 s_1 s_2 + \bar{s}^2 s_2^2 \, .
%\end{align}

%The $U(N+1|1)$ correction is
%\begin{align}
%    Z^+_{N+1|1} &= \frac{P^+_{N+1|1}}{(1-\bar{s} s_1)(1-\bar{s} s_2)} \, ,
%\end{align}
%with numerators
%\begin{table}[H]
%\begin{tabularx}{\columnwidth}{|c|X|}
%\hline
%$P_{1|1}^+$ & $\bar{s} s_1 + \bar{s} s_2 - \bar{s}^2 s_1 s_2$ \\ \hline
%$P_{2|1}^+$ & $\bar{s}^2 s_1^2 + \bar{s}^2 s_1 s_2 + \bar{s}^2 s_2^2 - \bar{s}^3 s_1^2 s_2  - \bar{s}^3 s_1 s_2^2$ \\ \hline
%$P_{3|1}^+$ & $\bar{s}^3 s_1^3 + \bar{s}^3 s_1^2 s_2 + \bar{s}^3 s_1 s_2^2 + \bar{s}^3 s_2^3 - \bar{s}^4 s_1^3 s_2 - \bar{s}^4 s_1^2 s_2^2 - \bar{s}^4 s_1 s_2^3$ \\ \hline
%\end{tabularx}
%\end{table}
%We confirm that the truncated expansion \eqref{eq:FbarF} holds and $Z_{N+2|2}^+=Z_{N+3|3}^+=0$ for $N=0,1,2$.

\subsection{$(\bar{F},F)=(2,1)$}

The $N=0,1,2$ partition functions are
\begin{align}
    Z_0 &= 1 \\
    Z_1 &= 1 + \bar{s}_1 s + \bar{s}_2 s \\
    Z_2 &= 1 + \bar{s}_1 s + \bar{s}_2 s + \bar{s}_1^2 s^2 + \bar{s}_1\bar{s}_2 s^2 + \bar{s}^2_2 s^2 \, .
\end{align}

The $U(N+1|1)$ correction takes the form
\begin{align}
    Z^+_{N+1|1} &= \frac{P^+_{N+1|1}}{(1-\bar{s}_1 s)(1-\bar{s}_2 s)} \, ,
\end{align}
with numerators
\begin{align}
P_{1|1}^+ &= \bar{s}_1 s + \bar{s}_2 s - \bar{s}_1\bar{s}_2 s^2 \\
P_{2|1}^+ &= \bar{s}^2_1 s^2 + \bar{s}_1\bar{s}_2 s^2 + \bar{s}^2_2 s^2 - \bar{s}_1^2\bar{s}_2 s^3  - \bar{s}_1\bar{s}_2^2 s^3 \\
P_{3|1}^+ &= \bar{s}_1^3 s^3 + \bar{s}^2_1\bar{s}_2 s^3 + \bar{s}_1\bar{s}_2^2 s^3 + \bar{s}_2^3 s^3 - \bar{s}^3_1\bar{s}_2 s^4 - \bar{s}^2_1\bar{s}^2_2 s^4 - \bar{s}_1\bar{s}_2^3 s^4 \, .
\end{align}

For $N=0,1,2$, we confirm that the truncated expansion holds,
\begin{align}
    Z_\infty = Z_N + Z_{N+1|1}^+ \, ,
\end{align}
and that
\begin{align}
    Z_{N+2|2}^+=Z_{N+3|3}^+=0 \, .
\end{align}

\subsection{$(\bar{F},F)=(2,2)$}

The $N=0,1,2$ partition functions are
\begin{align}
    Z_0 = 1 & \\
   Z_1 = 1 & + \bar{s}_1 s_1 + \bar{s}_1 s_2 + \bar{s}_2 s_1 + \bar{s}_2 s_2 + \bar{s}_1\bar{s}_2 s_1 s_2 \\
    Z_2 = 1 & + \bar{s}_1 s_1 + \bar{s}_2 s_1 + \bar{s}_1 s_2 + \bar{s}_2 s_2 + \bar{s}_1^2 s_1^2 + \bar{s}_1 \bar{s}_2 s_1^2 + \bar{s}_2^2 s_1^2 + \bar{s}_1^2 s_1 s_2 + 
    2 \bar{s}_1 \bar{s}_2 s_1 s_2 + \bar{s}_2^2 s_1 s_2 \nonumber \\
    &+ \bar{s}_1^2 s_2^2 + \bar{s}_1 \bar{s}_2 s_2^2 + \bar{s}_2^2 s_2^2 +\bar{s}_1^2 \bar{s}_2 s_1^2 s_2 + \bar{s}_1 \bar{s}_2^2 s_1^2 s_2 + \bar{s}_1^2 \bar{s}_2 s_1 s_2^2 + 
    \bar{s}_1 \bar{s}_2^2 s_1 s_2^2 + \bar{s}_1^2 \bar{s}_2^2 s_1^2 s_2^2 \, .
\end{align}

The two corrections are
\begin{align}
    Z^+_{N+k|k} &= \frac{P^+_{N+k|k}}{(1-\bar{s}_1 s_1)(1-\bar{s}_1 s_2)(1-\bar{s}_2 s_1)(1-\bar{s}_2 s_2)} \, , \quad k=1,2 \, ,
\end{align}
with numerators
\begin{align}
P_{1|1}^+ &= \bar{s}_1 s_1 + \bar{s}_2 s_1 + \bar{s}_1 s_2 + \bar{s}_2 s_2 -\bar{s}_1 \bar{s}_2 s_1^2 - \bar{s}_1^2 s_1 s_2 - 2 \bar{s}_1 \bar{s}_2 s_1 s_2 - \bar{s}_2^2 s_1 s_2 - 
    \bar{s}_1 \bar{s}_2 s_2^2 + \bar{s}_1^2 \bar{s}_2 s_1^2 s_2 + \bar{s}_1 \bar{s}_2^2 s_1^2 s_2 \nonumber \\
    & + \bar{s}_1^2 \bar{s}_2 s_1 s_2^2 + 
    \bar{s}_1 \bar{s}_2^2 s_1 s_2^2 - 2 \bar{s}_1^2 \bar{s}_2^2 s_1^2 s_2^2 \\ 
P_{2|1}^+ &= \bar{s}_1^2 s_1^2 + \bar{s}_1 \bar{s}_2 s_1^2 + \bar{s}_2^2 s_1^2 + \bar{s}_1^2 s_1 s_2 + \bar{s}_1 \bar{s}_2 s_1 s_2 + 
 \bar{s}_2^2 s_1 s_2 + \bar{s}_1^2 s_2^2 + \bar{s}_1 \bar{s}_2 s_2^2 + \bar{s}_2^2 s_2^2 - \bar{s}_1^2 \bar{s}_2 s_1^3 - \bar{s}_1 \bar{s}_2^2 s_1^3 \nonumber \\
 & - \bar{s}_1^3 s_1^2 s_2 - 2 \bar{s}_1^2 \bar{s}_2 s_1^2 s_2 - 2 \bar{s}_1 \bar{s}_2^2 s_1^2 s_2 -
  \bar{s}_2^3 s_1^2 s_2 - \bar{s}_1^3 s_1 s_2^2 - 2 \bar{s}_1^2 \bar{s}_2 s_1 s_2^2 - 2 \bar{s}_1 \bar{s}_2^2 s_1 s_2^2 - \bar{s}_2^3 s_1 s_2^2 - \bar{s}_1^2 \bar{s}_2 s_2^3\nonumber \\
 & - \bar{s}_1 \bar{s}_2^2 s_2^3 + 
 \bar{s}_1^3 \bar{s}_2 s_1^3 s_2 + \bar{s}_1^2 \bar{s}_2^2 s_1^3 s_2 + \bar{s}_1 \bar{s}_2^3 s_1^3 s_2 + 
 \bar{s}_1^3 \bar{s}_2 s_1^2 s_2^2 + \bar{s}_1^2 \bar{s}_2^2 s_1^2 s_2^2 + \bar{s}_1 \bar{s}_2^3 s_1^2 s_2^2 + 
 \bar{s}_1^3 \bar{s}_2 s_1 s_2^3 + \bar{s}_1^2 \bar{s}_2^2 s_1 s_2^3\nonumber \\
 & + \bar{s}_1 \bar{s}_2^3 s_1 s_2^3 - 
 2 \bar{s}_1^3 \bar{s}_2^3 s_1^3 s_2^3 \\
P_{3|1}^+ &= \bar{s}_1^3 s_1^3 + \bar{s}_1^2 \bar{s}_2 s_1^3 + \bar{s}_1 \bar{s}_2^2 s_1^3 + \bar{s}_2^3 s_1^3 + \bar{s}_1^3 s_1^2 s_2 + 
 \bar{s}_1^2 \bar{s}_2 s_1^2 s_2 + \bar{s}_1 \bar{s}_2^2 s_1^2 s_2 + \bar{s}_2^3 s_1^2 s_2 + \bar{s}_1^3 s_1 s_2^2+ \bar{s}_1^2 \bar{s}_2 s_1 s_2^2 \nonumber \\
 & + \bar{s}_1 \bar{s}_2^2 s_1 s_2^2 + \bar{s}_2^3 s_1 s_2^2 + \bar{s}_1^3 s_2^3 + \bar{s}_1^2 \bar{s}_2 s_2^3 + \bar{s}_1 \bar{s}_2^2 s_2^3 + \bar{s}_2^3 s_2^3 - \bar{s}_1^3 \bar{s}_2 s_1^4 - 
 \bar{s}_1^2 \bar{s}_2^2 s_1^4 - \bar{s}_1 \bar{s}_2^3 s_1^4 - \bar{s}_1^4 s_1^3 s_2 - 2 \bar{s}_1^3 \bar{s}_2 s_1^3 s_2\nonumber \\
 & - 2 \bar{s}_1^2 \bar{s}_2^2 s_1^3 s_2 - 2 \bar{s}_1 \bar{s}_2^3 s_1^3 s_2 - \bar{s}_2^4 s_1^3 s_2 - 
 \bar{s}_1^4 s_1^2 s_2^2 - 2 \bar{s}_1^3 \bar{s}_2 s_1^2 s_2^2 - 2 \bar{s}_1^2 \bar{s}_2^2 s_1^2 s_2^2 - 
 2 \bar{s}_1 \bar{s}_2^3 s_1^2 s_2^2 - \bar{s}_2^4 s_1^2 s_2^2 - \bar{s}_1^4 s_1 s_2^3  \nonumber \\
 &- 
 2 \bar{s}_1^3 \bar{s}_2 s_1 s_2^3 - 2 \bar{s}_1^2 \bar{s}_2^2 s_1 s_2^3 - 2 \bar{s}_1 \bar{s}_2^3 s_1 s_2^3 - 
 \bar{s}_2^4 s_1 s_2^3 - \bar{s}_1^3 \bar{s}_2 s_2^4 - \bar{s}_1^2 \bar{s}_2^2 s_2^4 - \bar{s}_1 \bar{s}_2^3 s_2^4 + 
 \bar{s}_1^4 \bar{s}_2 s_1^4 s_2 + \bar{s}_1^3 \bar{s}_2^2 s_1^4 s_2  \nonumber \\
 & + \bar{s}_1^2 \bar{s}_2^3 s_1^4 s_2 + 
 \bar{s}_1 \bar{s}_2^4 s_1^4 s_2 + \bar{s}_1^4 \bar{s}_2 s_1^3 s_2^2 + \bar{s}_1^3 \bar{s}_2^2 s_1^3 s_2^2 + 
 \bar{s}_1^2 \bar{s}_2^3 s_1^3 s_2^2 + \bar{s}_1 \bar{s}_2^4 s_1^3 s_2^2 + \bar{s}_1^4 \bar{s}_2 s_1^2 s_2^3 + 
 \bar{s}_1^3 \bar{s}_2^2 s_1^2 s_2^3 \nonumber \\
 &+ \bar{s}_1^2 \bar{s}_2^3 s_1^2 s_2^3 + \bar{s}_1 \bar{s}_2^4 s_1^2 s_2^3 + 
 \bar{s}_1^4 \bar{s}_2 s_1 s_2^4 + \bar{s}_1^3 \bar{s}_2^2 s_1 s_2^4 + \bar{s}_1^2 \bar{s}_2^3 s_1 s_2^4 + 
 \bar{s}_1 \bar{s}_2^4 s_1 s_2^4 - 2 \bar{s}_1^4 \bar{s}_2^4 s_1^4 s_2^4
\end{align}
and
\begin{align}
    P_{2|2}^+ &= \bar{s}_1^2\bar{s}_2^2 s_1^2 s_2^2 \\
    P_{3|2}^+ &= \bar{s}_1^3\bar{s}_2^3 s_1^3 s_2^3 \\
    P_{4|2}^+ &= \bar{s}_1^4\bar{s}_2^4 s_1^4 s_2^4 \, .
\end{align}

For $N=0,1,2$, we confirm that the truncated expansion holds,
\begin{align}
    Z_\infty = Z_N + Z_{N+1|1}^+ + Z_{N+2|2}^+ \, ,
\end{align}
and that
\begin{align}
    Z_{N+3|3}^+=0 \, .
\end{align}

%\subsection{$(\bar{F},F)=(3,2)$}

\end{document}